# List Sort

A New Approach for Sorting List to Reduce Execution Time

Adarsh Kumar Verma (*Student*)
Department of Computer Science and Engineering
Galgotias College of Engineering and Technology
Greater Noida, India
adarsh.verma1992@gmail.com

Prashant Kumar (*Student*)
Department of Computer Science and Engineering
Galgotias College of Engineering and Technology
Greater Noida, India
prashant.dwivedi@outlook.com

*Abstract*—In this paper we are proposing a new sorting algorithm, List Sort algorithm, is based on the dynamic memory allocation. In this research study we have also shown the comparison of various efficient sorting techniques with List sort. Due the dynamic nature of the List sort, it becomes much more fast than some conventional comparison sorting techniques and comparable to Quick Sort and Merge Sort. List sort takes the advantage of the data which is already sorted either in ascending order or in descending order.

## I. INTRODUCTION

Sorting is any process of arranging items in some sequence to reduce the cost of accessing the data. Sorting techniques can be divided in two categories, comparison sorting technique and non-comparison sorting techniques. A comparison sort is a type of sorting algorithms that only reads the list elements through a single abstract comparison operation that determines which of two elements should occur first in the final sorted list [1]. Comparison sorting includes Bubble sort, Insertion sort, Quick Sort etc[1]. Non-comparison sorting technique does not compare data elements to sort them. This category includes Bucket Sort, Count sort etc[1]. The efficient sorting algorithms which are based on comparison are Quick sort $O(n\log_2 n)$[7], Heap sort $O(n\log_2 n)$[11], Merge sort $O(n \log_2 n)$[5], Shell Sort $O(n(\log_2 n)^2)$[5] etc. Quick sort is found more advanced for inplace data which is initiated by C.A.R. Hoare in 1961[7] and it is considered as best sort algorithm for decades [5,6,8,10]. So far, for Quick sort, many variations in implementation[6,9] have been developed in order to overcome its drawbacks. Performance of the sorting algorithms is measured in the form of time and space complexities. Nowadays space complexity is not a big issue because memory getting cheaper and cheaper.

The proposed algorithm is a comparison sort based on the dynamic allocation of the items in the form of linked lists. Algorithm has two parts, insertion and merging. Since insertion is also comparison based, hence we are getting all the lists sorted. Insertion takes linear time to maintain the sorted list. When we reach at the maximum limit of the number of elements, we start merging the lists. If the data is already sorted in either ascending or descending order, we do not require merging and it sorts it within single pass. It has better execution time than basic sorting techniques like bubble, selection and insertion and is almost similar to quick and merge sorts.

## II. OVERVIEW AND ANALYSIS OF SOME WELL KNOWN SORTING TECHNIQUES

*Bubble Sort*

The bubble sort works by comparing each item in the list with the item next to it, and swapping them if required. The algorithm repeats this process until it makes a pass all the way through the list without swapping any item.

*Analysis:* Because of its simple and less complex nature bubble sort can prove effective when data to be sorted is very small. Performance of bubble sort in best case is $O(n)$ when list is already-sorted and it's worst case and average case complexity both are $O(n^2)$.

*Selection Sort*

This algorithm is called selection sort because it works by selecting a minimum/maximum element in each step of the sort. The number of passes, of the selection sort for a given list, is equal to the number of elements in that list [2]. The number of interchanges and assignments depends on the original order of the items in the list/array, but the sum of these operations does not exceed a factor of $n^2$ [3].

*Analysis:* It has $O(n^2)$ complexity, inefficient for the larger lists or arrays.

*Insertion Sort*

Insertion sort inserts each item into its proper place in the final list. It consumes one input at a time and growing a sorted list. It is highly efficient on small data and is very simple and easy to implement.

*Analysis:* It is highly efficient on small lists. The worst case complexity of insertion sort is $O(n^2)$. When data is already sorted it is the best case of insertion sort and it takes $O(n)$ time.

*Quick Sort*

Quick Sort is an algorithm based on the DIVIDE-AND-CONQUER paradigm that selects a pivot element and reorders the given list in such a way that all elements smaller to it are on one side and those bigger than it are on the other. Then the sub lists are recursively sorted until the list gets completely sorted.

*Analysis:* Quick sort is very efficient when data to be sorted is randomly scattered and it takes $O(n \log_2 n)$ time. And it does not perform well on nearly sorted data and give time complexity near about $O(n^2)$. Quick sort's performance is dependent on pivot selection. Most efficient versions of Quick Sort are not stable.

*Merge Sort*

This sorting method is an example of the DIVIDE-AND-CONQUER paradigm i.e. it breaks the data into two halves and then sorts the two half data sets recursively, and finally merges them to obtain the complete sorted list.

*Analysis:* Merge sort is very effective sorting technique when data size is considerably large. It Has the time complexity of _(n log n) for every case including worst, best and average case [4].

## III. WORKING PROCEDURE AND ALGORITHM OF LIST SORT

List sort algorithm has three parts which are calculation of maximum number of lists, insertion operation and merging operation. The procedure of sorting technique is to first insert the data into lists, and then merge these lists. The procedure of insertion and merging will continue until all the elements get sorted.

### A. Concept of Lists

Concept of lists represents that how many lists should be allocated to the data set to reduce time complexity. List is like a doubly ended queue in which data could be inserted at both ends. Fig. 1 shows the formation of the list, for the data set 10 there are 3 lists formed by insertion of the elements in the lists. Here it can be clearly depicted that all the lists are sorted itself, first elements of all the lists are sorted in ascending order and last elements of all the lists are sorted in descending order. The maximum number of the lists is first calculated by the list() function, which vary with the data range.

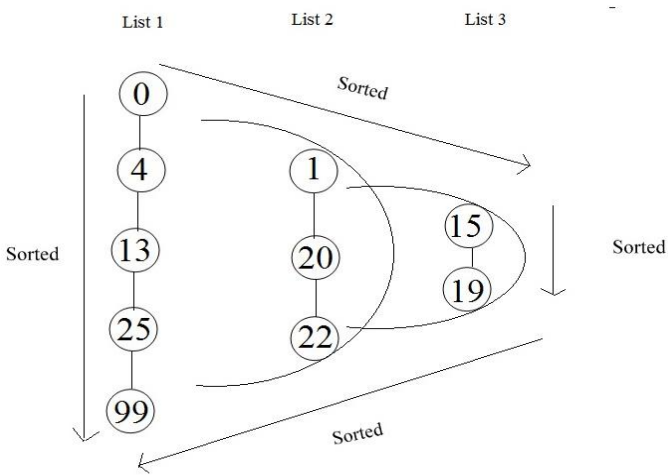

Fig. 1 Formation of List

We will provide total number of elements to be sorted (n) to list() function and it will return maximum no. of list (L), tt is number of ten thousands in the n.

```
list ( n )
    tt=n/10000
    if (tt<=0)
        return(15)
    else if (tt-10<=0)
        return(tt*5)
    else if (tt-50 <= 0)
        return(50+(tt-10)*3)
    else if (tt-100<=0)
        return(170+(tt-50)*1)
    else if (tt-500<=0)
        return(220+(tt-100)*0.45)
    else if (tt-1000<=0)
        return(400+(tt-500)*0.15)
    else if (tt-10000)
        return (475+(tt-1000)*0.032)
    else
        return ( 10000)
```

### B. Insertion

Insertion takes place in the list by comparing first and last element of the list, if new element is smaller than first element then insert it before the head and if new element is greater then last element of the list then insert it after last element otherwise create a new list and insert that element in to that list. Repeat the process up to L lists after that merge all the lists.

```
insert(data)

Node tmp;
insert the data into temp

// define current_list somehow before reaching this point or
make it available before calling this function by making
current_list as global

    if current_list.head is NULL
        assign current_list's head and tail point to tmp
            return
    else if data is less than current_list.head.data
            make temp current_list's head.
            return
    else  if data is greater than or equal to current_list.tail.data
            Assign tmp as current_list's tail
            return
```

else if we are at the last permissible node, then merging is required

 if current_list. position is equal to L
   //position starts from 1
      call merge(current_list)
 end if

// free tmp before new call to insert to avoid memory leak. Or you can simply assign tmp when inserting

 free(tmp);

 insert(data)

end function insert

### C. Merging of Lists

Merge() function tries to merge all the lists into one in a recursive fashion. It starts with the last list as current_list. Then current list is merged with its previous list, and so on until all the lists are merged into one single list.

Optimization: Out of current list and its previous list, we can merge the smaller list of the two into the bigger one. This reduces the number of insertions into the final list. In our program, this optimization is clearly visible when data set is large.

```
function merge(List current_list)
 List previous_list
 Node a, b, c, d
 //define previous list before beginning
 previous_list  = current_list.previous

 if current_list.position is not equal to 1
     Assign 'c' as current_list.head
     Assign 'a' as previous_list.head
     Assign 'b' as a.next
     Assign 'd' as c.next

     do while c is not NULL
        if c.data is less than b.data
           a.next <-- c
           d <-- c.next
           c.next <-- b
           a <-- c
           c <-- d
        end if
        else
           a <-- b
           b <-- b.next
        end else
     end while
     delete(current_list)
     previous_list.next <-- NULL
     //now we call merge again recursively with the previous list
     current_list <-- previous_list
     merge(current_list)
 end if
end function merge.
```

## IV. CASE STUDY FOR LIST SORT

In this section we are presenting the comparison of list sorting with other well known sorting techniques on various data sets and various orders of data.

### A. Best case

Best case of list sort occurs when data is sorted either in ascending or in descending order, because in this case time will be taken by only insertion operation, no merging operations are required. It takes O(n) time to sort n elements. From fig.2 below, we can see how sorting is done in the best case.

For ascending order

$T(n) = 1+ 1 + 1 + \ldots$ ( n times)

$T(n) = n$

### B. Average case

The average case of the List sort is evaluated on the random data and the time complexity evaluated is $(n*T/(L-1))$. Where n is the total number of elements, T is the total number of lists which are formed by the elements at run time and L is the number of list defined for the n elements. In random case it is found that T for 100000 elements is about 2000 and L is between 50 and 85.

For one time insertion, number of comparisons needed are

$S(n) = 1+2+\ldots+2L$

$S(n) = 2L(2L+1)/2$

Since the L list formed at T/L times, hence total insertion comparisons

$S(n) = L(2L+1)*T/L$

$S(n) = 2*T*L+T$

Hence for insertion, complexity is $O(T*L)$.

Now we will calculate time complexity for merging.

After insertion into L lists, merging function will be called. Let 'n' elements are divided in T list as n/T (taking for average case). For first set of L lists, the number of comparisons are:

$s(n) = (n/T-2) + (n/T-2) +\ldots$ ( L times, because there are L lists)

$s(n) = L*n/T - 2*L$

Since there are L lists formed T/(L-1) times hence we will calculate it for n elements

$S(n) = s(n)*T/(L-1) + s(n)*(T/(L-1) - 1) + s(n)*(T/(L-1) - 2) +\ldots+$ ( upto T/(L-1) times )

$S(n) = s(n)*\{ T/(L-1)*T/(L-1) - T/(L-1) \}$

$S(n) = s(n)*(T/(L-1))^2 - s(n)*T/(L-1)$

Substituting the value of s(n) and calculations

$S(n) = n*L*T/(L-1)^2$

Taking L/(L-1) approximately 1

$S(n) = n*T/(L-1)$

*C. Worst case*

The time complexity in the worst case will be $O((n^2)/(8*L))$, where n is the number of elements and L is the number of list needed at run time. List sort has advantage over other sorting techniques which gives $O(n^2)$ in worst case. The list sort will be in it's worst case when there is a fix pattern like { 1,10, 2, 9, 3, 8, 4, 7, 5, 6 }. Since the worst case is based on the fixed pattern, hence the probability of getting worst case is very low. Let T is the total no. of lists at run time and L is the maximum number of runtime usable lists defined by us. Hence total no. of times insertion and merging will be done is T/L. But in the worst case T will be n/2 because 2 elements in each list will be inserted, hence n/2 lists will be formed and n/2*L times insertion and then merging will be done.

Total no. of steps for insertion for the first time

$= 0+1+2+\ldots\ldots(2L \text{ times})$

$= 2*L^2 + L$

For inserting n/2*L times, total no. of insertions:

$T(n) = n/(2*L)*(2*L^2+L)$

$T(n) = n*L + n/2$

$T(n) = O(n*L)$

Now calculating the merging steps from last to second list

$= 2+4+6+8+10+\ldots$ ( upto (L-2) times)

$= L*(L-2)$

Total no. of merging steps for n/2*L times

= (merging from last to second list )+ (merging with first list)

Merging with first list for first time, no. of steps = 1, since there are only 2 elements and we need not to compare with first and last elements of first list because we have already compared them.

Hence total merging for first time

$= L*(L-2) + 1$

Total no of steps of merging for second time

$= L*(L-2) + (2*L)/2 + 2*(L-1)$

For simplification we can write as

$= L*(L-2) + (2*2-2)*(L-1)/2 + 2*(L-1)$

Total no. of steps of merging for third time

$= L*(L-2) + (2*3-2)*(L-1)/2 + 2*(L-1)$

Total no. of steps of merging for fourth time

$= L*(L-2) + (2*4-2)*(L-1)/2 + 2*(L-1)$

Total steps for merging = ( first time merging ) + ( second time merging ) + (third time merging )+…..(n/2L times)

$= n/(2*L)*(L*(L-2)) + n/(2*L)*2*(L-1) + ((L-1)/2)(2+4+6\ldots\text{upto } n/(2*L) \text{ times})$

$= n^2/8*L + n/4*L + n*L/2$

For simplification purpose L/(L-1) taken as 1.

Hence taking the dominating term $T(n) = O(n^2/(8*L))$.

## V. COMPARISON WITH WELL KNOWN SORTING TECHNIQUES

In the below graph, on the x-axis we have placed the number of elements and on the y-axis we have taken the time taken by the program for random number generator and execution in milliseconds. Execution time does not include printing of sorted result.

*Comparison with simple sorting techniques*

Here we are presenting comparison of the list sorting with the basic comparison sorts like bubble, selection, insertion sort. From the fig. 2 it is very clear that list sort is much more efficient than these sorting techniques including insertion sort, which is one of the best comparison sort.

TABLE I. COMPARISON OF TIME

| Number of Elements | Time Taken by Sorting Techniques ( in ms) | | | |
|---|---|---|---|---|
| | *Insertion Sort* | *Selection Sort* | *Bubble Sort* | *List Sort* |
| 5000 | 31 | 78 | 78 | 2 |
| 10000 | 125 | 296 | 328 | 5 |
| 20000 | 437 | 1140 | 1250 | 10 |
| 30000 | 1000 | 2672 | 2812 | 15 |
| 40000 | 1750 | 4703 | 4968 | 24 |
| 50000 | 2734 | 7188 | 7766 | 31 |
| 100000 | 10906 | 24719 | 30970 | 62 |

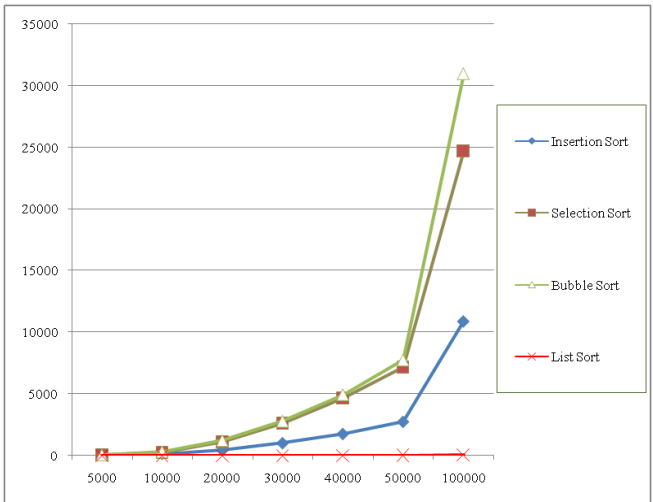

Fig 2 This is a Plot of Time Comparison of the Bubble Sort (GREEN), Selection Sort (BROWN), Insertion Sort (BLUE) and List sort (RED) over the Random Data Generated by

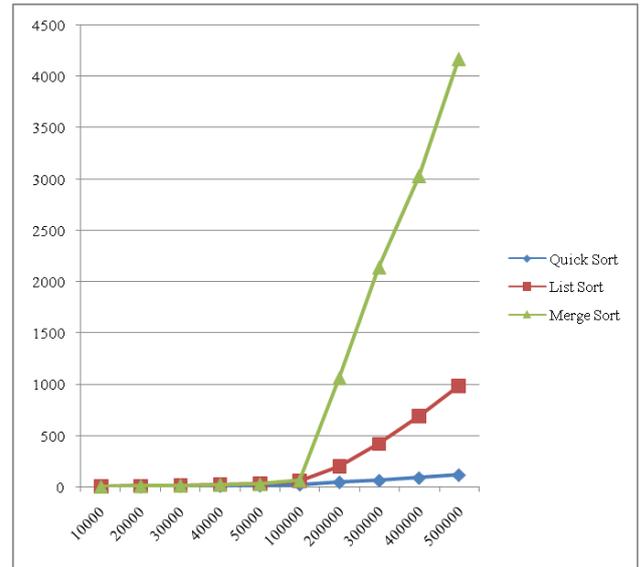

Fig. 3 This is a Plot of Time Comparison of the Quick Sort (BLUE), List Sort (BROWN) and Merge Sort (GREEN) over the Random Data Generated by Random Number Generator.

*Comparison with sorting techniques based on divide and conquer approach*

Here we are comparing List sort with two very well known and one of the best sorting techniques Merge sort and Quick Sort. Both of these sorting algorithms are based on Divide and Conquer paradigm. These results are evaluated by the array implementation of Quick and Merge sort. It is well known that Quick sort is typically faster than merge sort when the data is stored in memory, that's why Quick sort is leading in the array implementation. From the figure it is clearly depicted that List sort is also as efficient as the sorting techniques based on divide and conquer approach.

TABLE II.   COMPARISON OF TIME

| Number of Elements | Time Taken by Sorting Techniques ( in ms) | | |
|---|---|---|---|
| | *Quick Sort* | *List Sort* | *Merge Sort* |
| 10000 | 4 | 5 | 4 |
| 20000 | 6 | 10 | 9 |
| 30000 | 10 | 15 | 15 |
| 40000 | 11 | 24 | 25 |
| 50000 | 15 | 31 | 30 |
| 100000 | 25 | 62 | 62 |
| 200000 | 48 | 203 | 1062 |
| 300000 | 65 | 421 | 2140 |
| 400000 | 91 | 688 | 3031 |
| 500000 | 120 | 985 | 4172 |

*Comparison with merge sorts(linked list implementation)*

Here we are representing sorting by dynamic space allocation. We are considering the most two efficient sorts, Quick Sort and Merge Sort and comparing them with the List Sort. When the data set is huge and is stored on external devices such as a hard drive, Merge Sort is the clear winner with Quick Sort in terms of speed. It minimizes the expensive reads of the external drive and also lends itself well to parallel computing. It is well known that Quick Sort can't be efficiently implemented with the immutable data structures just like linked lists. But the linked list implementation of the List Sort is even more efficient than the Merge Sort. From the figure it is clearly depicted that List sort outperforms with the linked list implementation than any other sorting technique.

TABLE III.   COMPARISON OF TIME

| Number of Elements | Time Taken by Sorting Techniques ( in ms) | |
|---|---|---|
| | *List Sort* | *Merge Sort* |
| 5000 | 2 | 62 |
| 10000 | 5 | 265 |
| 20000 | 10 | 1218 |
| 30000 | 15 | 2969 |
| 40000 | 24 | 5485 |
| 50000 | 31 | 10876 |

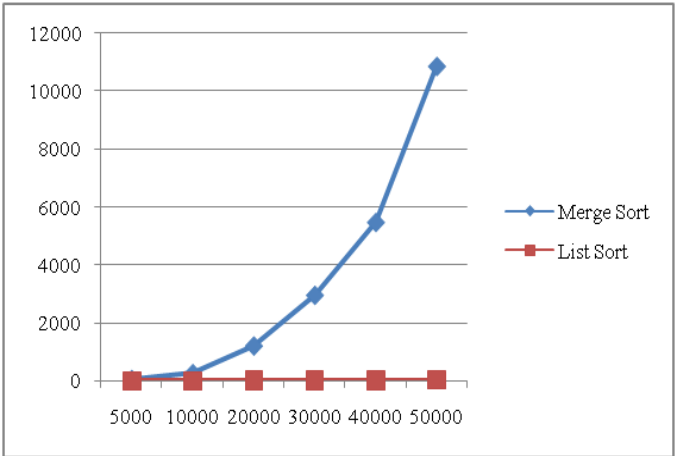

Fig. 4 This is a Plot of Time Comparison of the Merge Sort (BLUE) and List Sort (BROWN) (Linked list implementation) over the Random Data Generated by Random Number Generator.

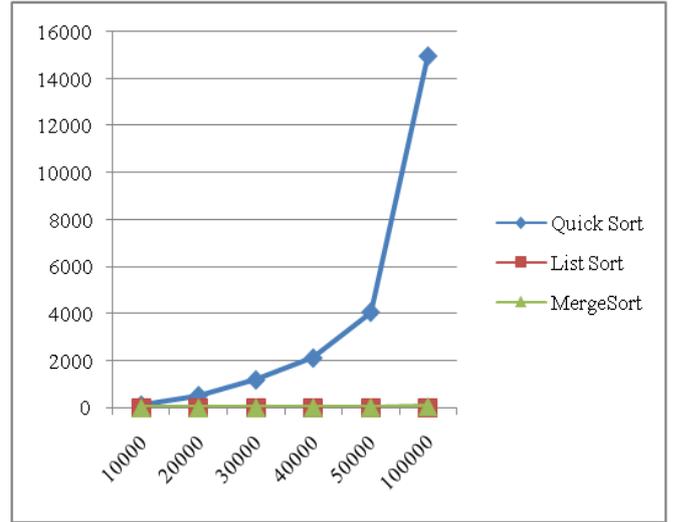

Fig. 5 This is a Plot of Time Comparison of the Quick sort (BLUE), List sort (BROWN) and Merge Sort (GREEN), when data is in either descending or in ascending order.

*Comparison with various sorting techniques with data either in ascending or in descending order*

Practically, most random large data sets contain sorted chunks of data in both ascending and descending orders. This is the worst case for the Quick Sort but Merge Sort will perform as usual. Since such an order is the best case for List Sort, it performs very well and gives results in linear time. From the below table and the figure it is very clear that performance of the List Sort in this case is outstanding.

TABLE IV. COMPARISON OF TIME

| Number of Elements | Time Taken by Sorting Techniques ( in ms) | | |
|---|---|---|---|
| | *Quick Sort* | *List Sort* | *Merge Sort* |
| 5000 | 78 | 0.05 | 2 |
| 10000 | 135 | 0.1 | 4 |
| 20000 | 535 | 0.5 | 9 |
| 30000 | 1200 | 1 | 15 |
| 40000 | 2130 | 1.5 | 25 |
| 50000 | 4077 | 2 | 30 |
| 100000 | 14953 | 5 | 62 |

## VI. CONCLUSION

Since nowadays data is sorted dynamically, List Sort will be an efficient choice for the dynamic allocation of space for elements to be sorted. From the comparison we can easily conclude that List Sort is far better than most of the comparison sorts and its dynamic storage allocation (linked list) implementation is better than Quick sort and Merge Sort. Another point to be noted here is worst cases of List Sort is hard to determine for large lists because it depends on a fix pattern which is practically rare to find.

## VII. FUTURE SCOPE

We would like to point on the fact that we have been unable to find a mathematical correlation between number of elements being sorted 'n' and the maximum number of lists 'L' which the algorithm should use to hold the elements before merging. While testing the algorithm, we found that this number varied from machine to machine being used for testing the algorithm.

As the value of L increases, execution time for sorting the same data set decreases, but after exceeding an 'optimum' value of L, execution time starts increasing. This optimum value, we think, depends on hardware configuration of the running computer and needs to be worked upon.

Due to lack of resources and time constraints, we are unable to look into this interesting feature of this algorithm. We hope that someone somewhere has the means and time to dig into it.

## VIII. ACKNOWLEDGMENT

We would like to express our gratitude to Divya Jaiswal, Anuj Tiwari (Students, Department of Computer Science & Engineering, GCET) And Navneet Kumar (Sudent, Department of Computer Science & Engineering, Sharda University) for their motivation and support.